\documentclass
[aps,prl,superscriptaddress,reprint,showpacs,amssymb,amsmath,floatfix,citeautoscript]{revtex4-1}

\usepackage{color}       
\usepackage{graphicx}                             
\graphicspath{ {./_Figures/} }
\usepackage{epstopdf}
\usepackage{color}
\definecolor{red}{rgb}{1.00,0.00,0.00}

\date{\today}

\begin{document}

\title{Direct observation of anapoles by neutron diffraction}

\author{S. W. Lovesey}
\affiliation{ISIS Facility, STFC, Didcot, Oxfordshire OX11 0QX, United Kingdom}
\affiliation{Diamond Light Source, Didcot, OX11~0DE, United Kingdom}

\author{T. Chatterji}
\affiliation{Institut Laue-Langevin, 71 Avenue des Martyrs, 38000 Grenoble, France}

\author{A. Stunault}
\affiliation{Institut Laue-Langevin, 71 Avenue des Martyrs, 38000 Grenoble, France}

\author{D. D. Khalyavin}
\affiliation{ISIS Facility, STFC, Didcot, Oxfordshire OX11 0QX, United Kingdom}

\author{G. \surname{van der Laan}}
\affiliation{Diamond Light Source, Didcot, OX11~0DE, United Kingdom}

\begin{abstract}

The scope of magnetic neutron scattering has been expanded by the observation of electronic Dirac dipoles (anapoles) that are polar (parity-odd) and magnetic (time-odd).    A zero-magnetization ferromagnet  Sm$_{0.976}$Gd$_{0.024}$Al$_2$ with a diamond-type structure presents Dirac multipoles at basis-forbidden reflections that include the standard (2, 2, 2) reflection. Magnetic amplitudes measured at four such reflections are in full accord with a structure factor calculated from the appropriate magnetic space group.
\end{abstract}


\date{\today}
\maketitle

Scattering experiments are omnipresent in scientific investigations, from the science of materials to the structure and function of biological systems to intrinsic properties of sub-atomic particles. Electronic properties of materials are revealed at an atomic level of detail by illuminating samples with beams of photons (wavelengths from the optical to hard x-ray regions of the spectrum), electrons, or neutrons, principally. Sample environments are routinely engineered to simulate extreme conditions found in the earth's crust or, for the case in hand, chosen so that quantum mechanics rules the sample's response to the illuminating radiation. Magnetic neutron scattering had an interesting birth in the 1930s when two theoreticians, both of whom later won a Nobel Prize, failed to agree on the form of the neutron-electron interaction \cite{Schwinger1937,Bloch1936,*Bloch1937}. Our experiments and calculations add a dimension that was not considered by them, namely, the parity-odd interaction that significantly expands the horizon of a technique established for parity-even scattering by magnetic electrons. In this context, we recall the ``totalitarian principle'' attributed to Murray Gell-Mann by which anything not forbidden (by symmetry) is compulsory \cite{Milton2006}.

Neutron Bragg diffraction has long been the method of choice in studies of magnetic materials for determining the configuration of magnetic dipoles \cite{Bertaut1963}, starting with Schwinger's correct theory of neutron-electron scattering in 1937 \cite{Schwinger1937} and a demonstration by Shull and Smart \cite{Shull1949} in 1949 of antiferromagnetic order in NaCl-type MnO below 122 K. Following in their footsteps, we advance compelling arguments in this Letter to show that neutrons are additionally scattered by anapoles, also known as toroidal dipole moments   \cite{Dubovik1990,Kopaev2009}. 
In presenting our case we unveil new features of a metallic zero-magnetization ferromagnet (ZMF) based on SmAl$_2$, lightly doped with Gd, which is ferromagnetic below a temperature $T_c$ $\approx$ 127 K \cite{Chatterji2018}.

Anapoles are magnetic (time-odd) and polar (parity-odd) dipoles that belong to a broader class of electronic Dirac multipoles characterized by the same discrete symmetries. Dirac multipoles are essential ingredients in some theories of ceramic, high-$T_c$ superconductors \cite{Varma2000,*Varma2006} and magnetoelectric materials \cite{Fiebig2005,Spaldin2008}. They have been observed using resonance-enhanced x-ray Bragg diffraction in a number of materials, e.g., vanadium sesquioxide (V$_2$O$_3$) and copper oxide (CuO) \cite{Lovesey2007,Scagnoli2011}. (Our Dirac multipoles describe static, equilibrium electronic properties not to be confused with objects of similar ilk ascribed to excitations in pyrochlore materials \cite{Morris2009}.) However, the interpretation of this type of x-ray diffraction is not always straightforward \cite{Joly2012,Lovesey2014-125504}, and a technique for the direct observation of Dirac multipoles will be highly prized. Already, neutron Bragg diffraction by the pseudo-gap phase of ceramic superconductors has been interpreted in terms of Dirac quadrupoles \cite{Lovesey2015-JPCM}. However, the observation of long-range magnetic order by neutron diffraction in these materials is seriously questioned \cite{Croft2017,Bourges2018} meaning that a robust demonstration of neutron diffraction by Dirac multipoles is left wanting. To this end, Bragg spots in the magnetic neutron diffraction pattern of the ZMF material Sm$_{0.976}$Gd$_{0.024}$Al$_2$ reported in this Letter, coupled to symmetry-informed calculations, provide a persuasive affirmative answer.

An interpretation of the diffraction pattern for the Sm magnetic compound of interest can sensibly start with basis-forbidden structural reflections in diffraction by diamond, first investigated by Bragg in 1921 \cite{Bragg1921}, and epitomized by the reflection indexed by Miller indices ($H_{\mathrm{o}}, K_{\mathrm{o}}, L_{\mathrm{o}}$) = (2, 2, 2). Electrons probed at these reflections occupy orbitals with opposite parities, because carbon sites in diamond are related by spatial inversion. Now, SmAl$_2$ possesses the diamond-type structure ${\mathit{Fd}}{\overline{3}}{\mathit{m}}$ (\#227, C15 cubic Laves structure) and neutron diffraction indexed by ($H_{\mathrm{o}} + K_{\mathrm{o}} + L_{\mathrm{o}}$) = $4n+2$, where $n$ is an integer is due to Al nuclei alone. In consequence, below the magnetic phase transition, $T \le T_c$, a magnetic contribution to a Bragg spot indexed by this condition must use unpaired electrons with opposite parities, e.g., Dirac multipoles created with Sm $4f$ and $5d$ electrons. The magnetic contribution in question is identified using polarization analysis in our experiments.

Measurements were performed on a large single crystal of nominal composition Sm$_{0.976}$Gd$_{0.024}$Al$_2$ that was used in Ref.~\cite{Chatterji2018}. Magnetization measurements confirmed its ZMF properties. Polarized neutron diffraction measurements were made on the spin-polarized diffractometer D3 on the hot neutron source at the Institut Laue-Langevin. The single crystal was mounted in an asymmetric split-pair cryomagnet and magnetized with a vertical field of 9 T parallel to the crystallographic [1, 1, 0] direction. The asymmetry in the peak intensities of Bragg reflections for 0.50 \AA\ neutrons polarized parallel and antiparallel to the field direction were made at temperatures in the range of $\sim$10-70 K. Significant asymmetry was measured in two allowed and in two forbidden reflections pairs.  

\begin{figure}[h]
\begin{center}
 \centerline{\includegraphics[width = 5.5cm]{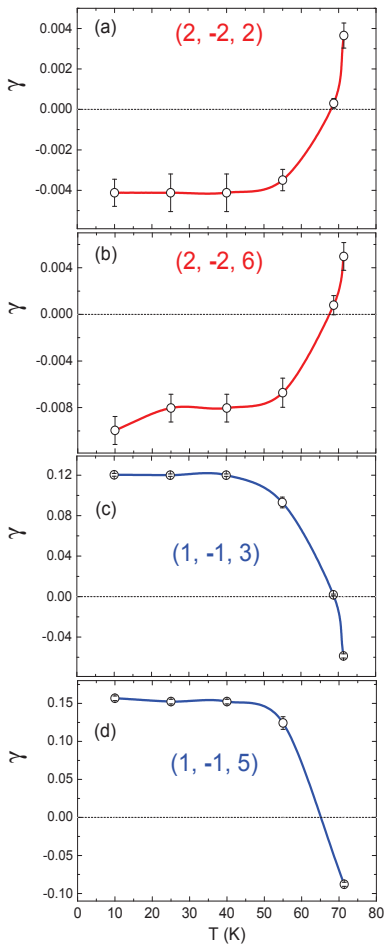}}
\caption{
Magnetic amplitudes as a function of temperature measured in an interval that embraces the compensation temperature, $T_{\mathrm{comp}}$ $\approx$ 68.7 K, of Sm$_{0.976}$Gd$_{0.024}$Al$_2$. Panels (a) and (b) 
show $\gamma = F^{(-)} / F_N$
for (2, $-$2, 2) and (2, $-$2, 6) basis-forbidden Bragg spots attributed to anapoles.  Displayed values are a simple average of measurements for ($H_{\mathrm{o}}$, $-H_{\mathrm{o}}$, $L_{\mathrm{o}}$) and $-$($H_{\mathrm{o}}$, $-H_{\mathrm{o}}$, $L_{\mathrm{o}}$) that are identical according to Eq.~(\ref{eq:2}). 
Panels (c) and (d) show $\gamma = F^{(+)}/F_N$  for (1, $-$1, 3) and (1, $-$1, 5) Bragg spots attributed to axial magnetic dipoles.
}
\label{fig:1}
\end{center}
\end{figure}

\begin{figure}
\begin{center}
\centerline{\includegraphics[width = 7.0cm]{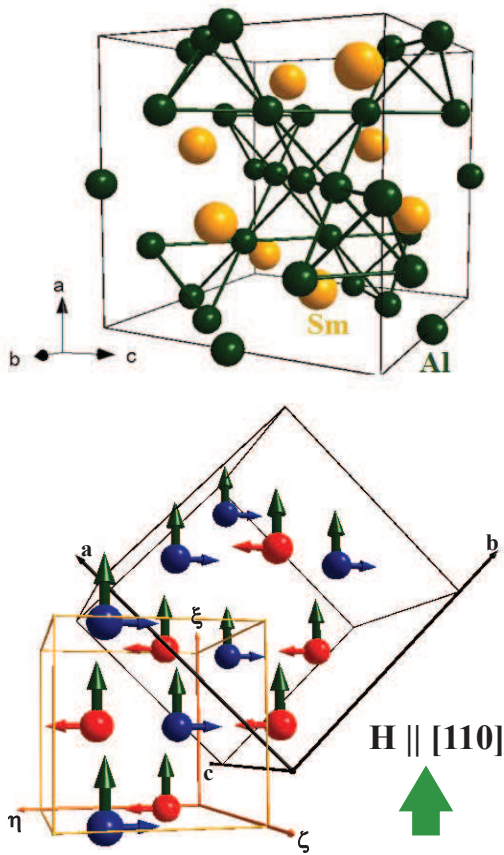}}
\caption{
Top panel; crystal structure (\#227,  C15 cubic Laves) with Sm ions in yellow and Al ions in green, and cell edges ($a$, $b$, $c$). Bottom panel; dipoles for magnetization parallel to [1, 1, 0] using a magnetic space-group ${\mathit{Imm'a'}}$ (\#74.559) \cite{BNS}. Cubic parent cell outlined in black, and orthorhombic magnetic cell ($\xi, \eta, \zeta$) with $\xi$ = (1/2, 1/2, 0),  $\eta$ = (1/2, $-$1/2, 0), and $\zeta$ = (0, 0, $-$1) outlined in yellow. Green arrows are axial dipoles parallel to the $\xi$-axis, while blue and red arrows that lie along the $\eta$-axis denote anapoles related by point inversion.}
\label{fig:2}
\end{center}
\end{figure}

Figures \ref{fig:1}(a) and \ref{fig:1}(b) show magnetic amplitudes of Bragg spots (2, $-$2, 2) and (2, $-$2, 6), corresponding to $n$ = 0 and $n$ = 1, observed in neutron diffraction by Sm$_{0.976}$Gd$_{0.024}$Al$_2$ through a range of temperatures that embraces the compensation temperature $T_{\mathrm{comp}}$ $\approx$ 68.7 K, at which (axial) Sm magnetic dipoles are zero \cite{Chatterji2018}. Our calculated magnetic unit-cell structure factors enable us to identify Sm anapoles as the primary source of magnetic diffraction. Displayed quantities are ratios of the magnetic and nuclear unit-cell structure factors that are derived from the ratio in intensities observed with neutrons polarized parallel and antiparallel to the applied magnetic field. Amplitudes of Bragg spots (1, $-$1, 3) and (1, $-$1, 5) that we attribute to axial magnetic dipoles, as before \cite{Chatterji2018}, are included in Figs.~\ref{fig:1}(c) and \ref{fig:1}(d). The chemical and inferred magnetic structures of SmAl$_2$ are depicted in Fig.~\ref{fig:2}.

\begin{figure}[h]
\begin{center}
 \centerline{\includegraphics[width = 9.0cm]{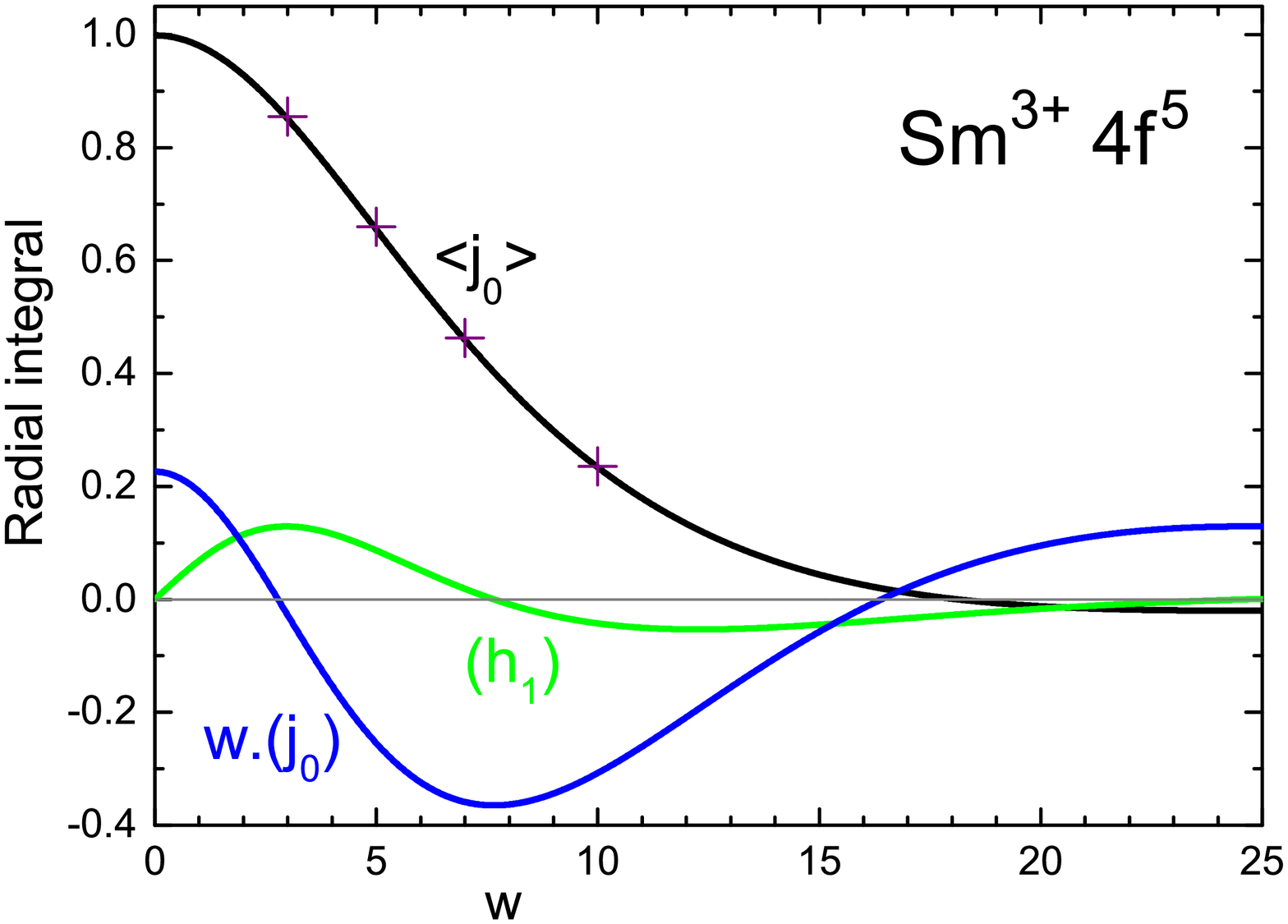}}
\caption{
Radial integrals for Dirac multipoles that appear in Eq.~(\ref{eq:2}) derived from an atomic code due to R. D. Cowan \cite{Cowan1968}. Dimensionless variable $w = 12 \pi a_{\mathrm{o}}s$, where $a_{\mathrm{o}}$ is the Bohr radius, while the standard variable for radial integrals $s$ is derived from the Bragg angle and neutron wavelength $s = \sin \theta / \lambda$. Green curve shows ($h_1$) and blue shows [$w \times (j_0)$]. Note that ($j_0$) is proportional to $1/w$ as the wavevector approaches zero. Atomic wavefunctions are $4f^5$--$5d^1$. Also included in the figure is the standard radial integral $\langle j_0 \rangle$   that appears in the so-called dipole-approximation (Eq.~\ref{eq:1}) for diffraction by axial dipole moments. Results obtained with our Sm$^{3+}$ ($4f^5$) wavefunction are denoted by the continuous black curve, to which we added for comparison four values (+) derived from the standard interpolation formula \cite{Brown2004}.
}
\label{fig:3}
\end{center}
\end{figure}

The plane of scattering gave access to the reflections ($-$$H_{\mathrm{o}}$, $H_{\mathrm{o}}$, $L_{\mathrm{o}}$). Samarium ions use sites $8a$ in \#227 with an origin (1/8, 1/8, 1/8) and cell length $a$ $\approx$ 7.943 \AA\ \cite{Chatterji2018}. 
Conditions ($H_{\mathrm{o}} + K_{\mathrm{o}}$), ($H_{\mathrm{o}} + L_{\mathrm{o}}$), and ($K_{\mathrm{o}} + L_{\mathrm{o}}$) even follow from $F$-centring. The Sm$^{3+}$ nominal atomic configuration is $4f^5$ ($^6H_{5/2}$) with a Land\'{e} $g$-factor $g$ = 2/7. Spontaneous magnetization parallel to [1, 1, 1] develops in SmAl$_2$ below a temperature $T_c$ $\approx$ 127 K. Orthorhombic ${\mathit{Imm'a'}}$ (\#74.559) \cite{Stokes,Campbell2006} is appropriate for magnetization parallel to [1, 1, 0]. Samarium ions are in sites $4e$ at an origin (0, 1/4, 1/8). Sites $4e$ possess symmetry ${\mathit{mm'}}2'$ \cite{BNS} that does not include spatial inversion, and admixtures of electronic orbitals with opposing parities are allowed. Absence of inversion symmetry at magnetic sites is a requirement for the existence of Dirac multipoles. The magnetic structure belongs to the crystal class $D_{2h}$ ($C_{2h}$) = ${\mathit{mm'm'}}$ that contains a centre of inversion symmetry, and a non-linear magnetoelectric effect is allowed.

The unit-cell structure factor for ${\mathit{Imm'a'}}$ contains 
$ [1 + \sigma_\pi \exp[-i \pi (H_{\mathrm{o}} + K_{\mathrm{o}} + L_{\mathrm{o}})/2] $, where $\sigma_\pi$ is the parity of the Sm time-odd multipole. The parity signature appears in the structure factor because the two Sm sites in the primitive cell differ by inversion. Axial magnetism is parity-even, $\sigma_\pi = +1$. Corresponding magnetic Bragg spots are absent for Miller indices ($H_{\mathrm{o}} + K_{\mathrm{o}} + L_{\mathrm{o}}$) = $4n+2$, and henceforth these are labelled basis-forbidden reflections. A dipole-approximation to the parity-even unit-cell structure factor is determined by the time-average of the Sm spin, $\langle \mathbf{S}_\xi \rangle$, and orbital moment, $\langle \mathbf{L}_\xi \rangle$, in the direction $\xi$ of the applied magnetic field, see Fig.~\ref{fig:2}. One finds
\begin{equation}
F^{(+)} \approx \frac{1}{2}  \left\{ 2 \langle {\mathbf{S}}_\xi \rangle \langle j_0 (k) \rangle + \langle {\mathbf{L}}_\xi \rangle 
[ \langle j_0(k) \rangle + \langle j_2(k) \rangle] \right\} ,
\label{eq:1}
\end{equation}
\noindent
where $k$ is the magnitude of the Bragg wavevector, and   $ \langle j_0(k) \rangle =  \langle 4f | j_0(kR) | 4f \rangle $ and $\langle j_2(k) \rangle$ are standard radial integrals with $\langle j_0(0) \rangle  = 1$ and $\langle j_2(0) \rangle  = 0$ \cite{Brown2004}.  Values of the quantities $2 \langle \mathbf{S}_{\xi} \rangle$ and $ \langle \mathbf{L}_{\xi} \rangle $ have been reported \cite{Chatterji2018}, and the magnetic moment 
$ \mu_{\mathrm{o}} = 2 \langle \mathbf{S}_{\xi} \rangle +  \langle \mathbf{L}_{\xi} \rangle $
 was found to vanish at a temperature $T_{\mathrm{comp}}$ $\approx$ 68.7 K. 
Quantities plotted in Figs.~\ref{fig:1}(c) and \ref{fig:1}(d) are $\gamma = F^{(+)}/F_N$   where the nuclear structure factor, $F_N$, is the sum of coherent scattering lengths for all elements in the sample.  
 
 Figures \ref{fig:1}(a) and \ref{fig:1}(b) contain our data for the ratio of magnetic and nuclear structure factors $F^{(-)}/F_N$ observed at basis-forbidden Bragg spots, and amplitudes attributed to multipoles that are both time-odd and parity-odd, $\sigma_\pi  = -1$. $F_N$ does not include the Sm coherent scattering length for these reflections. Diffraction patterns for  isostructural UAl$_2$ including basis-forbidden Bragg spots have been published \cite{Rakhecha1981}. However, the relatively poor quality of the data precludes a strong statement on the contribution of anapoles to magnetic Bragg diffraction \cite{Lovesey2018}, in contrast to the unequivocal statement that we are in a position to make on the basis of data in Fig.~\ref{fig:1}.

The calculation of Dirac multipoles for neutron scattering is complicated in the general case \cite{Lovesey2014-356001,Lovesey2015-PhysScr}. The monopole, or magnetic charge, is invisible in neutron scattering although it contributes in resonance-enhanced x-ray Bragg diffraction \cite{Staub2009}. The maximum rank of a Dirac multipole in neutron scattering is determined by the angular momenta of the atomic states. Diffraction by parity-even multipoles, and Eq.~(\ref{eq:1}) is the dominant part of the dipole, is similar, with even- and odd-rank multipoles up to an including a multipole of rank 7 in the case of rare-earth ions.

Dirac multipoles determine the actual value of $F^{(-)}$ according to magnitudes of radial integrals. Thus, we retain in $F^{(-)}$ multipoles accompanied by the largest radial integrals, in line with the construction of the dipole-approximation, Eq.~(\ref{eq:1}). The magnetic space-group allows Sm anapoles and the associated radial integrals are illustrated in Fig.~\ref{fig:3}. 
The anapoles in question are products of spin or orbital angular momentum with the electronic position operator, ${\mathbf{n}}$ \cite{Lovesey2014-356001,Lovesey2015-PhysScr}. We use  $ \langle \Omega  \rangle_{\mathrm{S}} = \langle {\mathbf{S}} \times × {\mathbf{n}} \rangle$ and 
$\langle \Omega \rangle_{\mathrm{L}} = \langle {\mathbf{L}} \times × {\mathbf{n}} - {\mathbf{n}} \times × {\mathbf{L}} \rangle$, and it is noted that operators ${\mathbf{S}}$ and ${\mathbf{n}}$ commute, whereas ${\mathbf{L}}$ and ${\mathbf{n}}$ do not commute. Anapoles are normal to the magnetic field in ${\mathit{Imm'a'}}$, as depicted in Fig.~\ref{fig:2}, and the basis-forbidden magnetic structure factor is     
 \begin{align}
F^{(-)} \approx &-i \exp [  \frac{ i \pi}{4} (H_{\mathrm{o}} - K_{\mathrm{o}}  - L_{\mathrm{o}} ) ]  \nonumber \\
& \times \kappa_\zeta  \, [3  \langle  \Omega_{\eta} \rangle_{\mathrm{S}} \, (h_1) -  \langle \Omega_\eta \rangle_{\mathrm{L}} \, (j_0) ] \, ,
\label{eq:2}
\end{align}
\noindent
with $ \kappa_\zeta = - L_{\mathrm{o}} /  \sqrt{ H^2_{\mathrm{o}} + K^2_{\mathrm{o}} + L^2_{\mathrm{o}} } $. Radial integrals ($h_1$) and ($j_0$) displayed in Fig.~\ref{fig:3} have been used to extract values of anapoles from our experimental data. 

According to Eq.~(\ref{eq:2}) the structure factor has equal values for partners in $\pm$($-$2, 2, $-$2) and in $\pm$($-$2, 2, $-$6), because a change in sign of $\kappa_\zeta \propto  L_{\mathrm{o}}$ is negated by a corresponding change in sign of the complex phase factor. Our observations are entirely consistent with this prediction, and values for $\gamma = F^{(-)}/F_N$  in Figs.~\ref{fig:1}(a) and \ref{fig:1}(b) are the average value of partners. Analysis of the data shows that both $\langle \Omega \rangle_{\mathrm{S}}$ and $\langle \Omega \rangle_{\mathrm{L}}$ change sign across the compensation temperature, and that these two anapoles have opposite signs. This statement assumes that the nuclear factor is the same in the two temperature regions 40 K and 72 K, to a good approximation. Aside from the nuclear structure factor; $\langle \Omega_\eta \rangle_{\mathrm{S}}$ $\approx$ 0.042 and $-$0.031 and $\langle \Omega_\eta \rangle_{\mathrm{L}}$ $\approx$ $-$0.155 and 0.092, for low and high temperature, respectively.

The direct observation by neutron scattering of Dirac (magnetoelectric) multipoles reported in this Letter extends the scope of neutron diffraction that is already the dominant technique for the determination of magnetic structures. In doing so, it opens a door to accurate results for Dirac multipoles that are fundamental entities in a raft of electronic properties \cite{Dubovik1990, Kopaev2009} that will complement and challenge simulations of electronic structures \cite{Fechner2016}. Materials include unconventional superconductors and those that display magnetoelectric effects, and a hidden-order using ruthenium Dirac multipoles proposed for metallic Ca$_3$Ru$_2$O$_7$ \cite{Thoele2018}.
A comprehensive theoretical analysis of our experimental data puts the key finding beyond reasonable doubt.

In summary, we presented neutron diffraction data and symmetry-informed calculations that attest to a direct observation of anapoles (Dirac dipoles) \cite{Milton2006,Lovesey2014-356001,Lovesey2015-PhysScr,Lovesey2017,Kopaev2009}. These parity-odd and time-odd multipoles should feature in future electronic structure calculations for the sample, Sm$_{0.976}$Gd$_{0.024}$Al$_{2}$, which is a well-characterized zero-magnetization ferromagnet \cite{Chatterji2018}.

\bibliography{ZMF-bib}

\end{document}